\documentclass[authoryear,preprint,review,12pt]{elsarticle}

\usepackage{graphicx}
\usepackage{amssymb}
\usepackage{makeidx}
\usepackage{bm}
\usepackage{epsfig}
\usepackage{psfrag}
\usepackage{lineno}
\usepackage{setspace}
\usepackage{amsmath}

\usepackage{amssymb}
\usepackage{natbib}
\journal{Physics of the Earth and Planetary Interiors}

\def\dsize{\displaystyle}

\begin{document}

\begin{frontmatter}



\title{Domino model in geodynamo}


\author[ph]{P.~Hejda}
\ead{ph@ig.cas.cz}
\author[mr]{M.~Reshetnyak\corref{cor1}}
\ead{m.reshetnyak@gmail.com}

\address[ph]{Institute of Geophysics, Academy of Sciences, 141 31
Prague, Czech Republic}

\address[mr]{
Institute of the Physics of the Earth, Russian Acad.~Sci,
 123995 Moscow, Russia}

\vskip 1cm \cortext[cor1]{ Corresponding author}

\begin{abstract}
  Using Lagrangian formalism we consider evolution of the ensemble of
 interacting  magnetohydrodynamic cyclones governed with  Langevin 
type equations in the rotating medium. This problem is relevant to
 the planetary cores where the Rossby numbers are small and geostrophic 
balance takes place. We show that variations of the heat flux at the outer 
boundary of the spherical shell modulates frequency of the reversals of the mean 
magnetic field that is in accordance with the 3D dynamo simulations. Two scenarios of reversals were observed. Either the axial dipole decreases in favour of quadrupole and then grows in opposite direction or the mean dipole tilts and reverses without decrease of its amplitude.
 \end{abstract}


\end{frontmatter}



\section{Introduction}
\label{s1}
The present geodynamo models include thermal and compositional convection which supply 
 the energy to the magnetic field generation in
 the rapidly rotating spherical shells~\citep{Tret}.
  Due to the strong non-linearity of the system and its three-dimensionality 
  the most reliable way of study was the numerical simulations. Because of the difference of the 
 convective and magnetic typical  times in order of magnitudes~\citep{Hol} simulation of the geomagnetic field
 evolution  is a very difficult and  time consuming problem even for the modern supercomputers. The analysis of the 
 obtained bulk of data is also very tricky process. This is motivation for considering simpler models,
  which can demonstrate some properties of the 3D models and present statistics for a longer time periods where 
  the complex models fail. This way helps not only to summarise some already known results but also to predict 
 directions of the 3D modeling, which has to deal with response of the model on various  interconnected parameters.

The choice of such a model should be based on the modern knowledge of the dynamo process in the core and be proved 
 by the observations. The specifics of the planetary convection is presence of the  geostrophic state, when 
  the scales along the axis of rotation are much larger than the scales in the perpendicular planes. The corresponding 
 anisotropy also takes place in the wave-space and leads to the inverse cascades even for the pure
 hydrodynamics~\citep{HR09}. 
  The  alongated along the axis of rotation 
   primitive convective cells are cyclones (anti-cyclones), in which the flowing up (down) liquid rotates 
 in such a way, that helicity of the  cell is  negative  in the northern hemisphere and positive 
 in the southern hemisphere. This set of cells produces  the mean magnetic field 
  at the surface of the planet, so that the cellular structure of the magnetic field is already smoothed and 
  the dipole configuration dominates. This very simple and naive approach was  successfully tested 
 in~\citet{NM12} where the net magnetic field was produced by the ensemble of the primitive spins
 located at the circle at the equatorial plane. The spins  rotate in the vertical plane,
  interact with each other and "feel" direction of the planet rotation. The energy is injected to the system 
 with the random force which mimics buoyancy sources. Such a system produces quite reasonable sequences of 
 the mean magnetic field reversals and can be adopted to the regimes with the variable frequency of the reversals.
 Here we want to extend this model, so that the spins could rotate in the horizontal plane as well. Transition to 
 rotation of the spins in the space makes possible to describe precession of the spins around the vertical axis 
  and, as a result, to mimic the magnetic pole wandering between the reversals and the fine structure 
 of the reversal itself. In fact, we observe  two different classes of reversals, when 
  the spins are synchronized in the horizontal plane during the reversal or not. The first class correspond to 
 the stable dipole field during the reversal and the second to the decaying field.  
 The other point is influence of the
 external fields on the system. Here we introduce analysis of the heat-flux modulation of the frequency reversals
  and compare our results with the 3D dynamo simulations. 
 We also consider how the core-mantle heterogeneity of the heat-flux can produce the preferred meridional band of
 the magnetic pole migration during the reversal.

\section{Two-dimensional approach}\label{BM1}
We briefly repeat results of \citet{NM12}, which is an extension of the Ising-Heisenberg 
XY-models of interacting magnetic spins. For more details of the history 
 of the problem and classification refer to~\citet{Stanley}. 
The main idea of the domino model is to consider a system of $N$
 interacting spins 
  ${\bf S}_i,\, i=1\dots N$, in media rotating  with angular velocity  ${\bm\Omega}=(0,\, 1)$ 
 in the Cartesian system of 
 coordinates $(x,\,y)$.
 The spins are located over an equatorial  ring, are of unit length and can vary angle $\theta$
  from the axis of rotation in the range of $[0,\, 2\,\pi]$  on  time $t$, so that  
${\bf S}_i=(\sin\theta_i,\, \cos\theta_i)$. 
 Each spin ${\bf S}_i$ is forced by a  random force, effective friction, as well as by  the 
closest  neighboring spins 
 ${\bf S}_{i-1}$ and ${\bf S}_{i+1}$. 

Following~\citet{NM12}, we introduce kinetic  $K$ and potential $U$ energies of the system:
\begin{equation}
\begin{array}{l}\dsize
K(t)={1\over 2}\sum\limits_{i=1}^N\dot{\theta}_i^2,     \\ \dsize
 U(t)=\gamma\sum\limits_{i=1}^N \left({\bm \Omega\cdot {\bf S}_i}
\right)^2+ 
\lambda \sum\limits_{n=1}^N 
\left(\bf{\bf S}_{i}\cdot {\bf S}_{i+1}\right).
\end{array}
\label{energy}
\end{equation}
The Lagrangian of the system then takes the form  ${\cal L}=K-U$. 
 Making the transition to the Lagrange equations,  
 adding friction proportional to
  $\dot{\theta}$ and the random force  $\chi$, 
\begin{equation}
{\partial \over \partial t}
{\partial{\cal L} \over \partial \dot{\theta}}
=
{\partial {\cal L} \over \partial \theta}-
\kappa\,\dot{\theta}+{\epsilon \chi\over \sqrt{\tau}},
\label{langevin}
\end{equation}
 leads to the system of Langevin-type equations:  
\begin{equation}
\begin{array}{l}\dsize
\ddot{\theta}_i-2\gamma\cos\theta_i\sin\theta_i+\lambda
\Big[
\cos\theta_i
\Big(
\sin\theta_{i-1}+\sin\theta_{i+1}
\Big)
-  \\ \dsize
\sin\theta_i
\Big(
\cos\theta_{i-1}+
\cos\theta_{i+1}
\Big)
\Big]  \dsize 
+\kappa\, \dot{\theta}_i+{\epsilon \chi_i\over \sqrt{\tau}}=0,\\
\dsize
  \theta_0=\theta_N,\, \theta_{N+1}=\theta_1,\, i=1\dots N,
\end{array}
\label{final}
\end{equation}
where $\gamma$, $\lambda$, $\kappa$, $\epsilon$, $\tau$ are constants.
 The measure of synchronization of the spins along the axis of rotation 
 \begin{equation}
\begin{array}{l}\dsize
M(t)={1\over N}\sum\limits_{i=1}^N\cos\theta_i(t)  
\end{array}
\label{Mdef}
\end{equation}
will be considered to be the total axial magnetic moment. 

With the appropriate 
 choice  of parameters the simulated sequences of $M(t)$  resembles paleomagnetic 
 records of the magnetic dipole evolution \\ \citep{NM12}, and
 possess some important properties of the geomagnetic field:
 has irregular time intervals between the reversals, exhibits
 short drops and recoveries of the field (the so-called excursions of the magnetic field), the 
 reconstructed 3D 
 magnetic field, based on idea that each spin is a magnetic dipole, has similar structure as the observable one. 
 
\section{Three-dimensional spin model}\label{BM20}
The natural next step is generalisation of the model to the 3D spins rotation. This  extension 
  helps to dispose 
 one of the critical disadvantages of the two-dimensional model. It is obvious, that nutation  displacements 
 in $\theta$-direction should cause response of  the Coriolis force perpendicular to $\theta$-
  and $\bm \Omega$-directions. 
 The Coriolis force causes precession of the spins around the axis of rotation 
 $\bm \Omega$  in the horizontal plane, like it happens for the magnetic spins in presence 
 of the external magnetic field or for the spinning top. 
 Formally this effect leads  to the new evolution equation for the azimuthal angle  
$\varphi$. Below we  consider how we can take into account this effect and compare it to 
  the known equations of the rotating bodies and magnetic spins. 

\subsection{Precession and  Landau-Lifshitz-Gilbert equation}
 In three dimensions direction of the spin 
 ${\bf S}_i=(\sin\theta_i\cos\varphi_i,\, \sin\theta_i\sin\varphi_i,\, \cos\theta_i)$ 
 is defined by two angles in the local spherical system of coordinates. 
 As before we suppose that origins of the coordinates are located equidistantly at the circle of 
 the unity radius at  the horizontal plane
 perpendicular to 
$\bm \Omega$. All the systems are obtained by translation in space, so that all corresponding axis are parallel.

First consider approximation when terms with 
  $\ddot\theta$, $\ddot\varphi$ and quadratic in $\dot\theta$ � $\dot\varphi$  are neglected.
 Then, using that fact that for the precession-like solution 
  the term proportional to $\dot\varphi\,\cos\theta$  should be included in the Lagrangian \citep{MAT02},
  this  leads to  the $i^{th}$ spin  
\begin{equation}
\begin{array}{l}\dsize
{\cal L}_i=
\dot{\varphi}_i\cos\theta_i -
\gamma \left({\bm \Omega\cdot {\bf S}_i}\right)^m -{\cal I}_i,
\end{array}
\label{lagr}
\end{equation}
 where   $m$ is integer. The interaction term can be written as follows:
\begin{equation}
\begin{array}{l}\dsize
{\cal I}_i=
\lambda\left[ 
\left({\bf S}_{i}\cdot {\bf S}_{i+1}\right) +
\left({\bf S}_{i}\cdot {\bf S}_{i-1}\right)
\right] = \\  \qquad \dsize\lambda \,
\Big[
\sin\theta_i 
\left( 
\sin\theta_{i+1} 
\cos(\varphi_i-\varphi_{i+1})
+
\sin\theta_{i-1} 
\cos(\varphi_i-\varphi_{i-1})
\right)
+
\\ \,\, \quad\qquad \dsize
\cos\theta_i \left(\cos\theta_{i+1}+\cos\theta_{i-1}\right)
\Big].
\end{array}
\label{Inter}
\end{equation}
 Further we consider two cases: 
i) $m=1$  corresponds to ferromagnetics, when direction of the external magnetic field is important and 
  the spins are co-directional to  $\bm \Omega$; 
 ii) for   $m=2$   the both directions $\pm \bm \Omega$ are equivalent, 
 so that the potential energy concerned with rotation is symmetric relative to the equatorial plane as it 
 was assumed in  2D case (\ref{final}). As according to
 the  paleomagnetic observations there is no preferable polarity 
 of the magnetic field and the Lorentz force in MHD equations is quadratic on the magnetic field,
  the latter case is
 more suitable for our tasks. As we see below the first two terms in r.h.s. of 
 (\ref{lagr}) for $m=1$ leads to Landau-Lifshitz-Gilbert (LLG) equation, which describes  precession 
 of the magnetic spin in the non-dissipative medium in the external Zeeman magnetic field equal to $\bm \Omega$.
 The third term describes local interactions of  the spins.
 Now we write the Lagrange equations for 4 independent variables
 $(\theta,\, \varphi,\, \dot\theta,\, \dot\varphi)$: 
\begin{equation}
\begin{array}{l}\dsize
{d \over d t}
{\partial{\cal L} \over \partial \dot{\theta_i}}
-
{\partial {\cal L} \over \partial \theta_i}+
{\partial {\cal F} \over \partial {\dot\theta_i}}
+{\partial {\cal R}_i\over \partial \theta}
=0, 
\\ \\ \dsize
{d \over d t}
{\partial{\cal L} \over \partial \dot{\varphi_i}}
-
{\partial {\cal L} \over \partial \varphi_i}
+
{\partial {\cal F} \over \partial \dot{\varphi_i}}
+{\partial {\cal R}_i\over \partial \varphi_i}
=0,
\end{array}
\label{lagr_eq}
\end{equation}
 where 
\begin{equation}
{\cal F}_i={\kappa\over 2}\,\left(\dot{\theta_i}^2+\sin^2\theta_i\,\dot{\varphi_i}^2\right),\qquad
{\cal R}_i=
{\epsilon \over \sqrt{\tau}}\,
\left(
{\theta_i}\,\chi_i \, +
 {\varphi_i}\, \psi_i 
 \right),
\label{Fdef}
\end{equation}
 and $\psi$ is the random function.

Substitution of  (\ref{lagr}) and (\ref{Fdef})  in (\ref{lagr_eq}) for $m=2$,  
 leads to the following system of equations:
\begin{equation}
\begin{array}{l}\dsize
\dot{\theta}_i - 
\kappa\, \sin\theta_i\,\dot{\varphi}_i
-
{{{\cal I}_i}'_\varphi \over \sin\theta_i}
-
{\epsilon\, \psi_i\over \sin\theta_i\,  \sqrt{\tau}}=0,\\
\dsize
\dot{\varphi}_i + 
\kappa\, {\dot{\theta}_i\over \sin\theta_i}
-
2\,\gamma\,\cos\theta_i +
{{{\cal I}_i}'_\theta \over \sin\theta_i}
+{\epsilon \chi_i\over \sin\theta_i\, \sqrt{\tau}}=0.
\end{array}
\label{final1}
\end{equation} 
 The exact form of the terms with derivatives of 
 ${\cal I}_i$ are given by equations:
\begin{equation}
\begin{array}{l}\dsize
-{{{\cal I}_i}'_\varphi  \over \sin\theta_i}=
\lambda\,\Big[
\sin\theta_{i+1}\,
\sin(\varphi_i-\varphi_{i+1})
+
\sin\theta_{i-1}\,
\sin(\varphi_i-\varphi_{i-1})
\Big],
\\  \\ \dsize 
{{{\cal I}_i}'_\theta \over \sin\theta_i}=
\lambda
\Big[
\cot\theta_i \Big( \sin\theta_{i+1}\, \cos(\varphi_i-\varphi_{i+1})
+ 
\sin\theta_{i-1}\, \cos(\varphi_i-\varphi_{i-1})\Big)
-  \\ \dsize \qquad \quad\quad \,\,\,\,
\Big(
\cos\theta_{i+1} +
\cos\theta_{i-1} 
\Big)
\Big].
\end{array}
\label{Int11}
\end{equation} 
Like before in 2D case (\ref{final}), we use the periodical boundary conditions: 
\begin{equation}
\begin{array}{l}\dsize
  \theta_0=\theta_N,\, \theta_{N+1}=\theta_1,\,\,
 \varphi_0=\varphi_N,\, \varphi_{N+1}=\varphi_1,\, i=1\dots N.
\end{array}
\label{bc1}
\end{equation}
 Since Lagrangian 
(\ref{lagr}) does not include quadratic terms proportional to 
$\dot\theta^2$,   $\dot\varphi^2$,
 equations 
(\ref{final1}) 
 present balance of the forces. 
 Before we come to analysis of the system 
(\ref{final1}), we consider case 
$m=1$, where  the term with   $2\,\gamma\,\cos\theta$  in (\ref{final1}) changes to  $\gamma$.
 Puting  $\lambda=0, \, \kappa=0,\, \epsilon=0$, yields 
 $\dot\theta=0$, $\dot\varphi=\gamma$, i.e. precession around axis $z$, that corresponds to 
 the solution of the LLG equation without dissipation for ferromagnetics:
\begin{equation}
\begin{array}{l}\dsize
{\dot  {\bf S}_i }=-\gamma \, {\bf S}_i\times {\bm \Omega},
 \end{array}
\label{llg}
\end{equation}
which has integral of motion 
$\dsize{\partial \,   \over \partial t} \,{\bf S}_i^2 = 0$,
 and for the constant in time 
$\bm \Omega$ the other integral: 
$\dsize{\partial  \over \partial t} \left({\bf S}_i\cdot {\bm \Omega} \right) = 0$, 
 that follows to 
$\dot\theta=0$. The  $x$-component of equation 
(\ref{llg}) gives the already mentioned above equation for the azimuthal angle:
$\dot\varphi=\gamma$.
  Existence of  precession distinguishes the 3D case from  2D, where
 the role of rotation comes to  attraction of the spins to the poles. 
(\ref{final}), so that  derivative 
  $\dsize {\partial \cal L}\over \dsize \partial \theta$
   moves from 
 $\theta$-component equation to  $\varphi$-equation.  

For the case 
$m=2$ one gets precession-like equation  $\dot\theta=0$, $\dot\varphi=2\,\gamma\,\cos\theta$, 
 which predicts reversal of the angular velocity of the spins at the equator plane.
  Obvious, that due to the distinguished direction concerned with rotation 
 the whole system is not symmetric to the change of $z\to -z$ and the break of the reflection symmetry for 
 $m=1$ gives preferable polarity of the magnetic field ($\overline{\sin\theta}\ne 0$).
  For  $m=2$  one has $\overline{\sin\theta}\to 0$, however $\dot\varphi$  changes its sign at the equatorial plane
  $z=0$. Below we consider this phenomenon in more details. 

  It is instructive  to consider
 balance of the curvilinear  terms  and  the Coriolis force  in    
$V_\theta$- and $V_\varphi$-components of 
 the Navier-Stokes equation in the spherical system of coordinates with $V_r=0$:
  \begin{equation}
\begin{array}{l}\dsize
-V_\varphi^2 \cot\theta=H\, V_\varphi\,\cos\theta\,, \qquad  
V_\theta \, V_\varphi \cot\theta=-H\,V_\theta\, \cos\theta, 
 \end{array}
\label{NS1}
\end{equation}
where $H$ is the amplitude of the Coriolis force,
 and tangential velocity 
 $(V_\theta,\, V_\varphi)$
 is 
$(\dot\theta,\, \dot\varphi\,\sin\theta)$.
 The both equations in 
  (\ref{NS1}) gives 
  $\dot\varphi=-H$, 
 that corresponds to the case with 
$m=1$ and $V_\theta$ remains undefined.
 As we see, condition of  equiprobability of the direct and inverse polarity for 
$m=2$ changes the whole scenario.

Return to 
(\ref{final1}) with $m=2$.
 Since we have we omitted quadratic in velocities terms, this approximation is valid for
 the slow regimes with precession between the reversals of the field.
  Anyway we  consider how the predicted asymmetry of the precession velocity is influenced by the 
 random force. Even for the small random force distribution of ${V_\varphi}(M)$ in Fig.\ref{fig1}a
  for the regime without reversals in  Fig.\ref{fig1}b has zero mean value. The reason
 of such a  seeming discrepancy  with our estimate $\dot\varphi\sim \cos\theta$ is the following.
 Consider averaged in time coupled equations (\ref{final1}) in the
 simplified  form without interaction, diffusion and random forces for small $\theta\ll 1$
 and  constant in time 
   between the reversals of the magnetic field
   $\overline{\dot\theta}=0$ and check if $\dot \varphi$ still changes sign for the coupled system:
  $\overline{\dot\theta}\sim \overline{\theta \dot\varphi}=0$, 
 $\overline{\theta \dot\varphi} +\overline{\dot\theta}\sim\gamma\overline{(1-\theta^2)\theta}$
  that leads  to  the contradiction:
$\overline{\theta \dot\varphi} \sim\gamma\overline{\theta}\ne 0$, 
 that means that the other terms should be included in the consideration and naive prediction of
 change of the sign for $\dot\varphi$ is wrong. 
 We conclude, that the random forcing makes system symmetric to the change of $\varphi$-direction.
 Here we do not consider extremely small $\epsilon$ regimes to preserve the realistic for the geomagnetic field
 level of the field fluctuations between the reversals. 
 
 The considered  approximation with the small fluctuations of $\theta$
  is reliable for some astronomical problems without nutation, 
 as well as to the ferromagnetic 
 problems when the temperature is less than the  Curie point and $\theta$-variations are small. 
 This review  was used from the methodological point of view and now we come to the full equations
 applicable to the highly non-linear regimes with the large accelerations.

\subsection{Precession and nutations}\label{BM3}
 Now we consider approximation  with large    $\ddot\theta$, $\ddot\varphi$, 
 $\dot\theta^2$, $\dot\varphi^2$ 
  using 
 analogy to the spinning top with unity moments of inertia \citep{Landau} with  Lagrangian in the form:
\begin{equation}
\begin{array}{l}\dsize
{\cal L}_i={1\over 2}\left(\dot\theta_i^2+\sin^2\theta_i\, \dot\varphi^2_i + (\dot\varphi_i 
\cos\theta_i+ \dot\zeta_i)^2   \right) -  
\gamma \left({\bm \Omega\cdot {\bf S}_i}\right)^2   -{\cal I}_i
 + {\Psi}_i
, \end{array}
\label{lagr1}
\end{equation}
 where  $\dot\zeta_i$ is angular velocity of the top.
 In our case 
 $\dot\zeta$ is fixed and given property of the spin.
 For simplicity we take it equal to unity: $\dot\zeta_i^2\equiv |{\bf S}_i|^2=1$.   The new variable 
 $\Psi$ is the potential  used latter for description of the external forces.

Finaly Lagrangian can be written as follows
\begin{equation}
\begin{array}{l}\dsize
{\cal L}_i={1\over 2} \,\dot\theta_i^2+{1\over 2} \, \dot\varphi^2_i +
\dot\varphi_i \,
\cos\theta_i -  
\gamma\, \cos^2\theta_i    -{\cal I}_i + {\Psi}_i,
 \end{array}
\label{lagr1a}
\end{equation}
 where constant is omitted because  Lagrange equations includes only derivatives of  
 $\cal L$. Neglecting of the quadratic terms in velocities in  (\ref{lagr1a})  leads to 
 the simplified version of
 Lagrangian  (\ref{lagr}). 

Substitution of 
 (\ref{lagr1}) in 
 (\ref{lagr_eq}) yields to the dynamic equations:
\begin{equation}
\begin{array}{l}\dsize
\ddot{\theta}_i  
+
\dot{\varphi}_i\,\sin\theta_i - 
\gamma\,\sin 2\theta_i +
{{\cal I}'_i}_\theta
+\kappa\, {\dot{\theta}_i}+{\epsilon \chi_i\over \sqrt{\tau}}
- {{\Psi'}_i}_\theta
=0,
\\ \dsize
\ddot\varphi-\sin\theta_i\, \dot{\theta}_i
+
{{\cal I}'_i}_\varphi
+
\kappa\, \sin^2\theta_i\,\dot{\varphi}_i+{\epsilon \psi_i\over  \sqrt{\tau}}
- {{\Psi'}_i}_\varphi
=0.\\
\end{array}
\label{final2}
\end{equation} 
 Neglecting the second in time derivatives and quadratic terms leads to the original system 
(\ref{final1}).

Note, that in contrast to 2D case we already have two equations, that requires "energy" balance fulfilment, 
 and the new numerical methods should be used. This point is enforced by that fact, that 
 there are terms with  $\sin\theta$ in the denominator. That was the reason to use Newton-Raphson iterative method 
 in the residual form described in the Appendix.

Now we consider important for the geomagnetism question on the predictability of the geomagnetic field reversal on
 the secular variations before the reversal \citep{Jacobs} and take as a measure of variations the amplitude 
 of the magnetic dipole precession velocity $ V_\varphi$ around the geographical pole. 
 Using equations (\ref{final2}) we generate 30 reversals of the field, see Fig.\ref{fig2}(1).
 Taking the typical time $t_i$ between the reversals as 
$3\, 10^{5}$y we come to the estimate of the full time interval as 
 $9\, 10^6$y and the time unity 
 $\dsize \tau_u= 3\, 10^{5}/ 2\, 10^4=4\, 500$y. 
 The estimate of the azimuthal velocity  $V_\varphi=0.2$
 leads in the order of magnitude to the typical archeomagnetic 
time estimate of the pole wandering (similar to the west drift velocity)
$\dsize \tau_a={\pi\over 8} \, {1\over V_\varphi}\tau_u\sim 3\, 500$y, where we 
 took into account, that between the reversals the pole locates in the cone 
 $\dsize \theta<{\pi\over 8}$.

 We inspect the set of the five reversals of the magnetic field, see Fig.\ref{fig2}(2).
  The particular feature of this regime is existence of the intermediate state corresponding to the small 
  $|M|$ when the system does not know in what direction it should go, and as a result some of the reversals 
 fails producing excursions,  see Fig.\ref{fig2}(1).
 To check possibility of the reversal predictability we consider the mean values of 
$\overline{M}$ and $|\overline{V_\varphi}|$ over 
  these five reversals, see Fig.\ref{fig2}(3, 4). As we see,  the change of polarity does not lead to the substantial 
 change of  
$|\overline{V_\varphi}|$,
 that means that the behaviour of the model is symmetric in respect to the moment of the reversal.

\section{Magnetic field of  the spins}
 The essential disadvantage of the domino model based on the assumption of the 
  long-range interactions  for the spins  
 is absence of connection between the spin's position and its magnetic field.  
 That was the reason why we still used integral characteristic  $M$ as a measure of the field.
  Let us summarise requirements to the model which we want to satisfy:
   i) the potential energy of the spins should have minimum when they are located near the poles and the spins 
 should have the same direction;
 ii)
 the model provides calculation of 3D magnetic field at least at some distance from the spins.
  Identification of the spins with the magnetic dipoles helped \cite{NM12} to get 3D  
 distribution of the  magnetic field for the spins in the form:
  \begin{equation}
\begin{array}{l}\dsize
{\bf B}={3\, {\bf r}\, ({ {\bf d}\cdot {\bf r}}) - r^2\, {\bf d} \over r^5},
 \end{array}
\label{D1}
\end{equation}
where  $\bf d$ is a magnetic moment of the dipole in the centre of the coordinates and
$\bf r$ is the radius-vector of observation. It should be noted, that 
as follows from (\ref{D1}), the stable state  of two magnetic dipoles at the distance  $R$
  with corresponding potential energy of interaction 
 \begin{equation}
\begin{array}{l}\dsize
U={3\, ({{\bf d}_1\cdot {\bf R}}) ({{\bf d}_2\cdot {\bf R}}) - ({{\bf d}_1\cdot {\bf d}_2})\, R^2 \over R^5}
 \end{array}
\label{D2}
\end{equation}
 is quadrupole, when the magnetic moments ${\bf d}_1$, ${\bf d}_2$  are anti-parallel.
  In its turn  this  would correspond to the death of the dipole magnetic field at large $R$ 
 and relation (\ref{D2}) is out of interest for  geomagnetism. 

To overcome this problem  and to get the self-organised mean dipole field \cite{NM12} 
 suggested to use the  simplified version of (\ref{D2}) with omitted first term and  opposite 
 sign for the second term, see 
 (\ref{energy}). This form is similar to that one in the ferromagnetics where magnetic domains enforce the external 
 magnetic field choosing the same with it  direction. In ferromagnetics  this effect has  the pure
  quantum origin concerned with the  exchange energy of the magnetic domains.
  As regards to  \citep{NM12} it was just postulated as a convenient way to get similar 
 to observations results and we join this idea in the rest of the paper.

Using equation  (\ref{D1}) for the set of  $N$ magnetic dipoles,
 located at the circle of unity radius we compute the vector magnetic field, see distribution of its 
  radial component 
 at the distance of the three unity length $\cal R$  during one of the reversal
  in Fig.\ref{fig3}. One of the specific features of this reversal is existence of the preferred
 meridional  band where the reversal  occures, see Fig.\ref{fig3}(2).
 To consider this effect in more details we come to analysis of the spectral properties, see
 evolution of the axial  dipole Gauss coefficient  $g_1^0$, in Fig.\ref{fig4}(1),
 at the  distance  $\cal R$ for the five reversals in Fig.\ref{fig2}(1).
 We conclude that the typical time of the drop and recover of the field during the
  reversal in this model is the same,  there is some increases of the field just before and
  after the reversal. So as the amplitude of the spins is fixed the net magnetic flux of $|{\bf B}|$
 is constant in time as well as its spectrum, during, e.g., the reversal.
 This leads to the exchange of the magnetic energy between the dipole and quadrupole modes in  
 Fig.\ref{fig4}(2), where the ratio of the dipole and quadrupole modes gives 
  $\overline{\cal D}/\overline{\cal Q}\approx 13$ compared in order of magnitudes with that at the
  surface of the Earth. In the moment of the reversal $t\sim 2\,100$   $\cal D$ drops,
 and intensity of the qudrupole $\cal Q$ increases to the level of the dipole field. This effect is known 
 in 3D simulations as well. The reason of such a distribution is that magnetic energy in the geostrophic systems 
 is quite large and hardly can be immediately transferred to the kinetic energy during the reversal. It means that 
  the drop of the dipole field should be compensated with increase of the higher modes.

The other, may be the more specific feature of the domino model is that for the other four 
 reversals  
  the amplitude of the total dipole field during  reversal is constant.  
  In contrast to the previous case, where due to the uncorrelated horizontal projections 
 of the  dipole field  the averaged 
 dipole field decreased, the second case corresponds to the coherent state when even during the 
 reversal  all the 
 spins have the same direction. For this scenario the mean  dipole field rotates in the meridional plane without
  decrease of the amplitude. As we see, this three-dimensional phenomenon   for the considered regime 
  is more expected and caused by 
 interaction  of the spins (\ref{final2}) in 
 $\varphi$-direction during the reversal with  the short  interaction 
 time compared to the times of the reversal and precession.
  The both scenarios does not contradict to observations.

\section{Thermal flux heterogeneities}
 One of the important results of the geodynamo theory is that frequency of the reversals depend on 
 the spatial distribution of the heat flux at the core-mantle boundary 
 \citep{GCHR99}.  Fluctuations of the heat flux of order 10--20\%  of the mean flux value
  caused with the processes in the mantle and D" layer have typical time scales  
$10^6$-$10^7$y, are 
 much larger time scales in the liquid core  $10^4$-$10^5$y. 
 In particular increase of the heat flux towards the poles is equivalent to increase of rotation, that damps 
 the reversals. Such fluctuations appears to be the thermal traps for reversals. 
 In its turn, the decrease of the heat flux in the high latitudes leads to chaotic behaviour of the magnetic dipole and 
 increase of the reversal frequency, when the
  Archimedean forces become more significant. 
  This effect corresponds to increase of the Rossby number, which has threshold 
at $0.12$ where regime of the stable dipole field changes to the regime of  the frequent reversals.
  It looks tempting to reproduce this effect in the domino model, where the large
 number of  the reversals can be easily simulated.

We now extend the concept of the spin from the purely magnetic system  to the whole cyclone system, 
 including its hydrodynamics, and we introduce correction   $\Psi(\theta,\, \varphi)$  to the
  potential energy 
  $U$, which 
 takes into account the  heterogeneity of the thermal flux. 
  The new effective force  $\dsize {\bf F}_i$ is proportional to the corresponding derivatives of 
 ${\Psi_i}$ with the opposite sign will appear in  (\ref{final2}).
 Here we study  influence of the various forms of $\Psi$ on  the behaviour of  $M(t)$.

   Let  $\Psi(t,\, \theta,\, \varphi)=C_\psi \, \psi(t,\, \theta,\, \varphi)$,  where $C_\psi$ is a constant,
 and the spatial distribution of the potential is given by $\psi=-\cos^2\theta$. 
  Accordingly to the recent estimates of the heat-flux variations at the core-mantle boundary,
 which can be about 20\%~\citep{Olson10}, we conclude that 
 $\dsize C_\psi\sim 0.2{ \epsilon\over  \sqrt{\tau}}\sim 1$. 
   Then, 
$C_\psi>0$ 
 corresponds to the stable state in  the polar regions,  
 $\theta=0,\, \pi$, and 
 the appearing force  
$F=-\sin 2\theta$, acting 
  on the cyclones, is directed towards the poles. This regime corresponds 
to the increase of the thermal flux near the poles
 that causes the stretching of the cyclone along the axis of rotation. 
 So as $F$ changes sign at the pole $\theta=\pi$, and the force is directed to the pole (or outward) for 
$\theta\to \pi-0$ and $\theta\to \pi+0$,  we can consider it as the spherical coordinate when it is needed.
   In Fig.\ref{fig5} 
  we demonstrate the effective influence of $F$ on  $M$ for the regime in \ref{fig2}(1).
 The increase of the thermal flux along the axis of rotation leads to the partial  suppression of the reversals
  of the field, Fig.\ref{fig5}(1). 
 Note that, for the chosen potential barrier $\psi$, the
  dependence of  $F$  is equal to  the 
 $\gamma$-term 
  in  (\ref{final}): 
the  increase of the thermal flux at the poles leads to the effective increase of rotation and amplification 
of geostrophy,
 caused by the  rapid daily rotation of the planet.
  Our results are in agreement with the 3D simulations, see Fig.1d in~\citep{GCHR99}.  
   The further increase of $C_\psi$ ($C_\psi=2$) leads to the total stop of the reversals.
 It is more interesting that, 
 using even larger $C_\psi> 10$, one arrives at regimes with a nearly 
 constant in time $|M|\le 1$ defined by the initial 
 distribution of ${\bf S}_i$. In other  words, the super flux at the poles can fix the spins which are still not coherent. 
  There is some evidence~\citep{Shats05} that the geomagnetic dipole in the past could have migrated from 
  the usual position near the geographic poles to some stable state in the middle latitudes. Within the framework of our model,
 we can explain this phenomenon by the thermal super flux at the poles. Later 
we will discuss some other scenarios which yield similar results.

For negative  $C_\psi$,
 when the geostrophy breaks due to the relative intensification of convection in the equatorial plane,
 we get the opposite result, see Fig.\ref{fig5}(2): the regime of the frequent reversals observed in
 Fig.1c in~\citep{GCHR99}. 
  In this  case force 
 $F$  is directed from the poles and the equilibrium point at  the poles becomes unstable. 
  The new minimum of the potential energy at the equator leads to the appearance of a new attractor, with the dipole at
 the low latitudes.  Similar behaviour of the magnetic dipole is observed on 
  Neptune and Uranus;  for more details see,  e.g.~\citep{CHR02}.

Now we consider the non-axial-symmetrical potentials. Let  $\psi = \cos^2\theta\cdot \cos^2\varphi$.
 It is clear to see, that for the large $|C_\psi|$ 
 frequency of the reversals should follow the previous axial-symmetrical  case 
 of $\psi$, in particular for  $C_\psi=3$ 
 the number of reversals decreases, see Fig.\ref{fig5}(3).
  Moreover, this form of $\psi$ with dependence on  $\varphi$
 leads to the new effect:
 appearing of the preferred meridional band of the magnetic poles migration during the reversals.
  We compare  distribution of  $h_1^1(g_1^1)$ without 
 potential, see Fig.\ref{fig6}(1), 
 with that one with the non-axialsymmetrical potential.
 For the first one we have isotropic distribution of the horizontal dipoles, see Fig.\ref{fig6}(1).
  Insertion of the potential leads to  appearing of the barriers for the reversals with  gaps at 
  $\varphi=0$ and $\varphi=\pi$, see 
  Fig.\ref{fig6}(2),
 where the large circles corresponding
 to the magnetic dipole at the equator plane have the elliptical distribution, and
 near the poles (small circles) distributed isotropicaly. 
 For the considered form of 
$\psi$ sign of $F_\theta$ did not change in 
$\varphi$. 

The last example  is with $F_\theta(\varphi)$ changing sign at the equator plane: 
 $\psi = \cos^2\theta\cdot \sin 2\varphi$, see Fig.\ref{fig6}(3), with the only one  reversals for the model
 simulation and the anisotropy observed for all positions of the integral dipole, even for the locations 
 near the poles, when the  amplitude of horizontal dipole 
 $\sqrt{g_1^{1\, 2}+ h_1^{1\, 2}}$ is small.
 In other words, fluctuations of the heat flux causes not only preferred meridional bands of the
 magnetic poles migration, but  the anisotropy of the poles locations between the reversals as well. 
 We also observe states, when for the large $|C_\psi|$ reversals stop because some of the spins
  come to the 
  thermal trap near the poles (for some particular values of $\varphi$) and the full reversal  which includes 
 reversals of all spins is impossible.

 \section{Conclusion}\label{concl}
  It appears, that even a small extension of the original toy domino model leads to some nice physical effects known
 in the various fields of physics.  In spite of simplicity of this approach it has very strong
  advantage: one can test 
 very different scenarios of magnetic field evolution basing on some realistic information on the cyclone 
 convection in the core. 
 This approach is well designed for modelling  of interaction on the different scales and number of the neighbour
 cyclones, which can be obtained from the 3D models. The other possibility is to use two distinct sets of  cyclones
 inside and outside of the Taylor cylinder, which separates two different  regions in the liquid core. 
  
 We also believe, that domino models can
  be modified for the analysis of the net kinetic and may be current, magnetic helicities as well.
   The first kind of helicities is the most preferable for the analysis, because the cyclone's  
 hydrodynamics  to some level is stable during the reversals, what is not the case for
 the generated in the cyclone magnetic  field. 
 The latter obstacle makes interpretation of the results in the domino model not so unambiguous.


\appendix

\section{}
\label{Ap}
We rewrite system  (\ref{final2}) in the form of the first order differential equations:
\begin{equation}
\begin{array}{l}\dsize
V_i-\dot\theta_i=0,\\ \dsize
W_i-\dot\varphi_i=0,\\ \dsize
\dot V  + \kappa\, V   +
W\sin\theta_i+
 A=0, \\ \dsize 
\dot W
-\sin\theta_i\, V+\kappa\,\sin^2\theta_i\, W+B=0,
\end{array}
\label{lin101}
\end{equation}
relative to the vector
\begin{equation}
{\bf y}^n=\left(
\begin{array}{l}\dsize
V^n\\ W^n \\
\theta^n\\ \varphi^n
\end{array}
\right),
\label{vect1}
\end{equation}
where 
\begin{equation}
\begin{array}{l}\dsize
A=-\gamma\,\sin 2\theta_i+
{{\cal I}'_i}_\theta+{\epsilon\,\chi_i\over\sqrt{\tau}}-
{{ \Psi}'_i}_\theta
,\qquad
B=
{{\cal I}'_i}_\varphi+{\epsilon\,\psi_i\over\sqrt{\tau}}-
{1\over \sin\theta_i}{{ \Psi}'_i}_\varphi.
\end{array}
\label{lin1011}
\end{equation}
Using the implicit first order Euler scheme in time
 for the $n^{th}$ time step we introduce residual vectors 
for (\ref{lin101}):
 \begin{equation}
{\bf e}^n=
\left(
\begin{array}{l}\dsize
V^n\, dt-\theta^n + \theta^{n-1}\\  \dsize
W^n\, dt-\varphi^n + \varphi^{n-1}\\ \dsize 
V^n-V^{n-1}+\kappa\,V^n\,dt
 +
W^n\sin\theta^n_i\,dt+
 A^n\,dt
\\ \dsize
W^n-W^{n-1}
-\sin\theta_i^n\,V^n\,dt +\kappa\,\sin^2{\theta_i}^n\,W^n\,dt+B^n\,dt
\end{array}
\right).
\label{lin22}
\end{equation}
The iterative  Newton-Raphson   process for the $p^{th}$-iteration 
 gives:
 \begin{equation}\dsize
{\bf y}_{p}^n={\bf y}_{p-1}^n-\left(
{\partial {\bf e}_{p-1}^n \over \partial {\bf y}^n_{p-1}}
\right)^{-1}\cdot {\bf e}_{p-1}^n,
\label{lin5}
\end{equation}
where   Jacobian matrix has form:
\begin{equation}
\begin{array}{l}\dsize
\hat J={\partial {\bf e}_p^n \over \partial {\bf y}^n_p}=
\begin{bmatrix}  \dsize
dt & 0 & -1 & 0 \\ \dsize
0 & dt & 0 & -1 \\ \dsize
1+\kappa\,dt  & 
\sin\theta_i^n\,dt
  & \dsize
W^n\cos\theta_i^n\,dt+{\partial A^n\over\partial \theta}\,dt
 & 
\dsize {\partial A^n\over\partial \varphi^n}\,dt \\ \dsize
-\sin\theta_i^n\, dt & 
1+
\kappa\,\sin^2\theta_i^n\, dt
 & J_{4\,3}
 & \dsize 
{\partial B^n\over\partial \varphi^n}\,dt
\end{bmatrix},
\end{array}
\end{equation}
��� 
\begin{equation}
\begin{array}{l}\dsize
J_{4\,3}=
-\cos\theta_i^n\,V^n\,dt+\kappa\sin2\theta_i^n\,W^n\,dt+
{\partial B^n\over\partial \theta^n}\,dt.
\end{array}
\label{Jac}
\end{equation}
For each time moment 
 $n$ and spin 
 $i$ one has iterative process 
(\ref{lin5}) with updated other  spin values.
 After the desired convergence is reached transition to  the new time step  $n+1$ is done.
 At the each time step we check conditions 
 $\theta_i\in (0,\, 2\,\pi)$, $\varphi_i\in (0,\, 2\,\pi)$ and correct values if necessary.
  This algorithm appears to be stable to singularity at the axis and can be easily transformed to the second-order 
 Crank-Nicolson scheme in time. Anyway this problem is more stable than the PDE in the spherical coordinates, because
 the large derivatives near the pole forces spins to go out from the pole region and instability 
 in such a self-organised way is damped. The spins are squeezed from the pole regions.

\begin{singlespace}

\end{singlespace}

\newpage
\pagestyle{empty}

{\Huge
\begin{figure}[th]
\psfrag{t}{ $t$}
\psfrag{M}{$M$}
\psfrag{Vf}{$V_\varphi$}
\psfrag{1a}{\Large 1}
\hskip 0.5cm \epsfig{figure=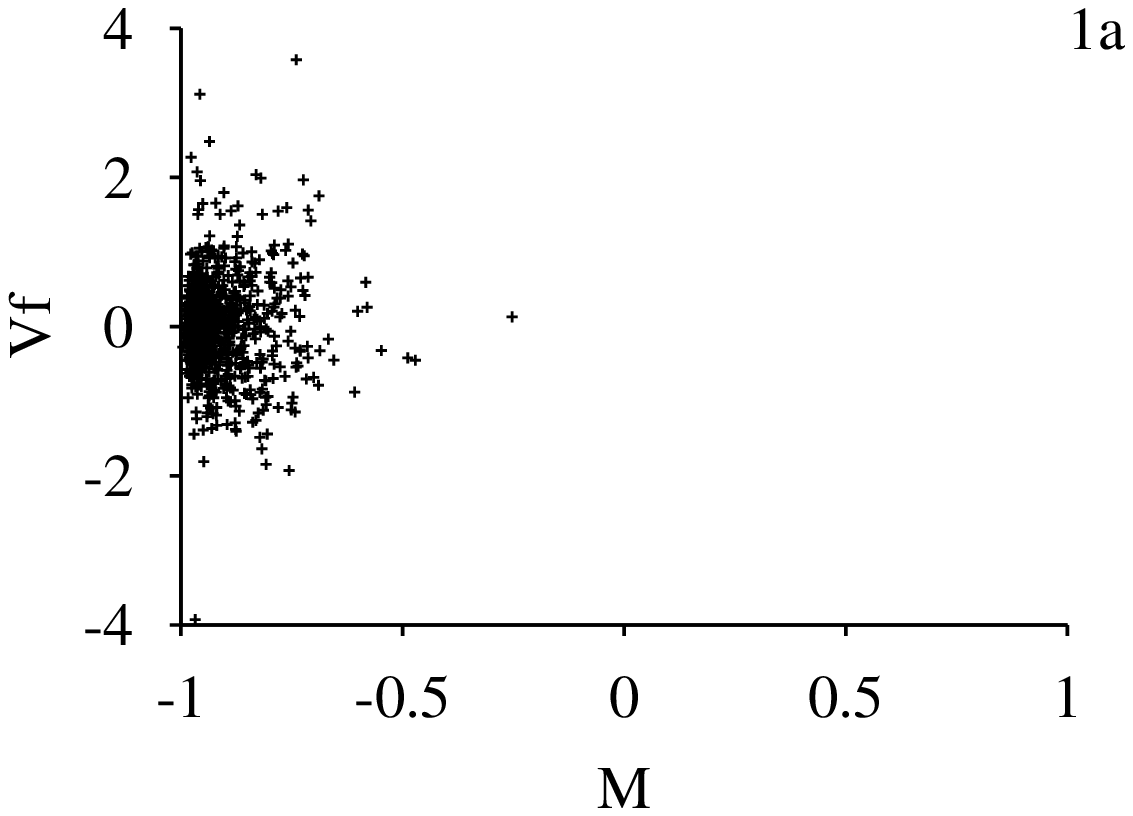,width=9cm}
\vskip 0.1cm
\psfrag{1a}{\Large 2}
\hskip 0.25cm \epsfig{figure=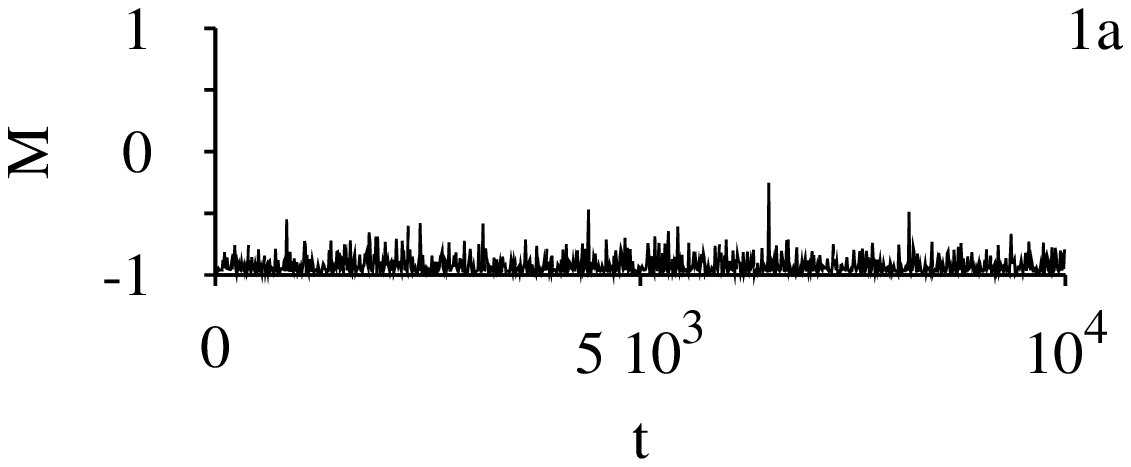,width=9.3cm}
%
\vskip 0.4cm
 \caption{
 Dependence of velocity precession 
 $V_\varphi$ on dipole polarity 
$M$ (1) and  evolution of the dipole  $M$ in time (2) 
 for $\gamma=-1.8$, $\lambda=-6$, $\kappa=0.1$, $\epsilon=0.03$, $\tau=0.01$.  
 }
\label{fig1}
\end{figure}
} 

\eject\newpage
{\Huge
\begin{figure}[th]
\vskip -2.5cm
\psfrag{t}{ $t$}
\psfrag{M}{$M$}
\psfrag{1a}{\Large 1}
\hskip 0.5cm \epsfig{figure=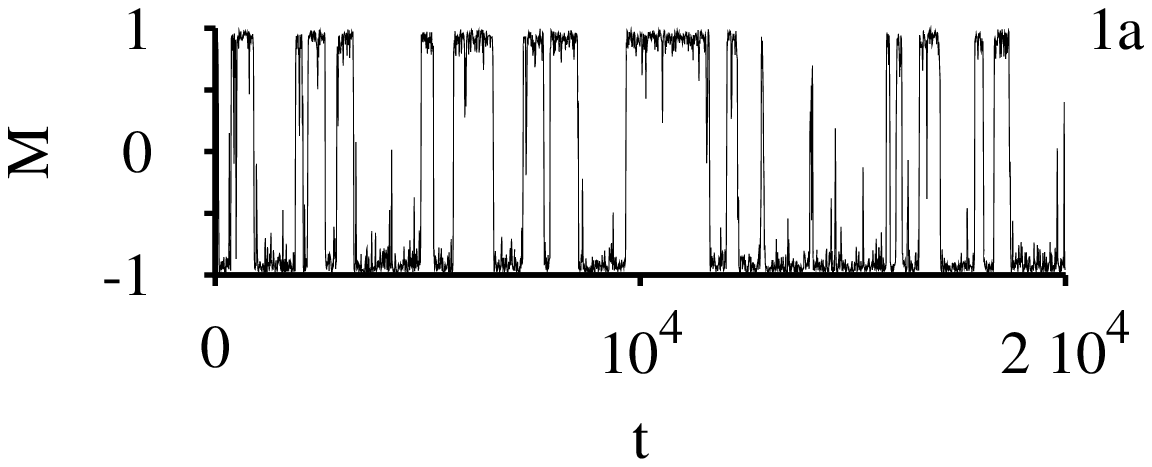,width=9cm}
\vskip -0.5cm
\psfrag{1a}{\Large 2}
\hskip 0.5cm \epsfig{figure=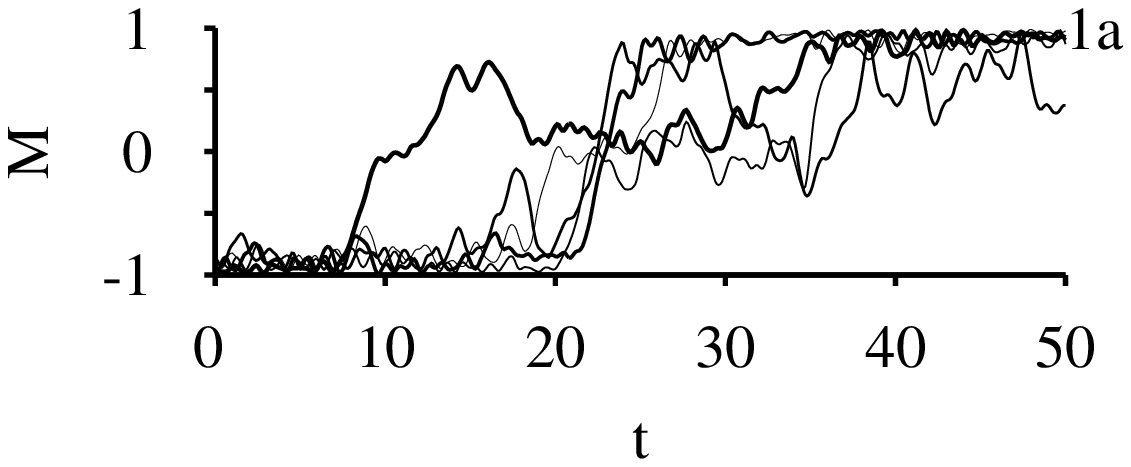,width=9cm}
\vskip -0.5cm
\psfrag{1a}{\Large 3}
\psfrag{M}{$\overline{M}$}
\hskip 0.5cm \epsfig{figure=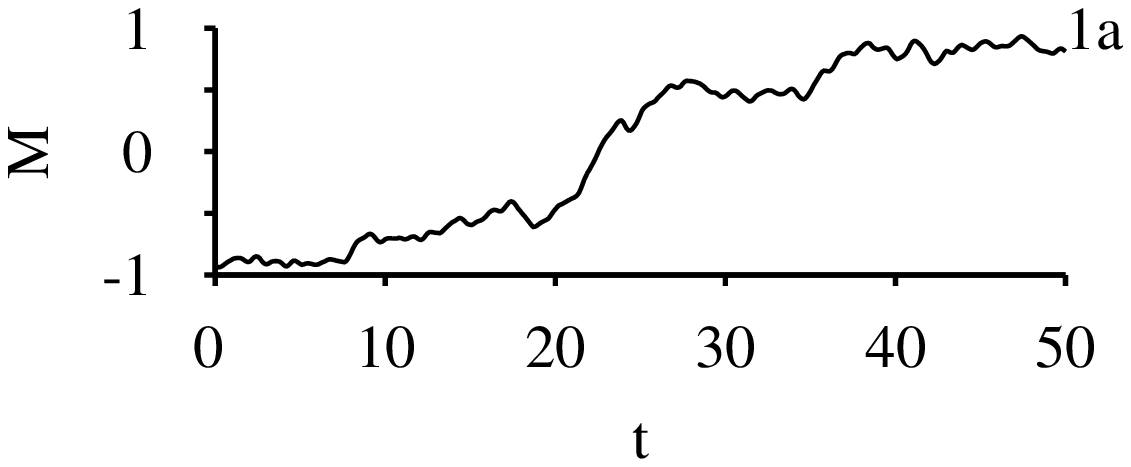,width=9cm}
\vskip -0.5cm
\psfrag{1a}{\Large 4}
\psfrag{Vf}{$|\overline{V_\varphi}|$}
\hskip 0.5cm \epsfig{figure=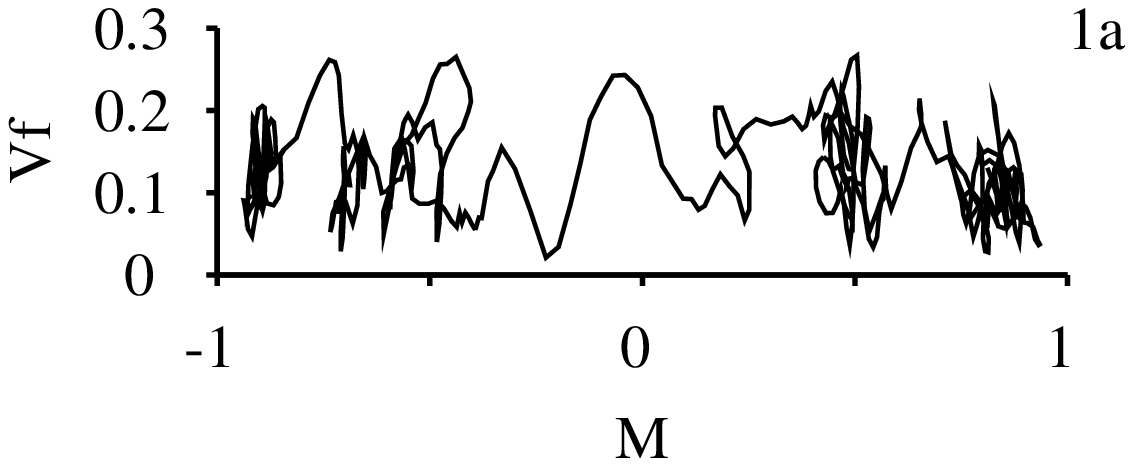,width=9cm}
\vskip 0.1cm
%
\vskip 0.4cm
 \caption{ 
 Evolution in time of  $M$ for  $\gamma=1$, $\lambda=-4.8$, $\epsilon=0.8$, $\kappa=0.2$ (1) and
 reversals of the magnetic field (on degree of thickness of the line) for 5 time intervals 
$(9640,\, 9690)$, $(12000,\, 12050)$,  $(12260,\, 12310)$, $(17840,\, 17890)$,  $(18670,\, 18720)$
 where for the 3rd and 5th interval sign of $M$ is changed  (2); (3) -- the mean value 
 $\overline{M}$ for the previous  plot (2)  and the mean precession velocity 
  $|\overline{V_\varphi}|$ vice  $\overline{M}$ (4).
}
\label{fig2}
\end{figure}
}

\eject\newpage

{\Huge
\begin{figure}[th]
\vskip -2.5cm
\hskip 1.5cm \epsfig{figure=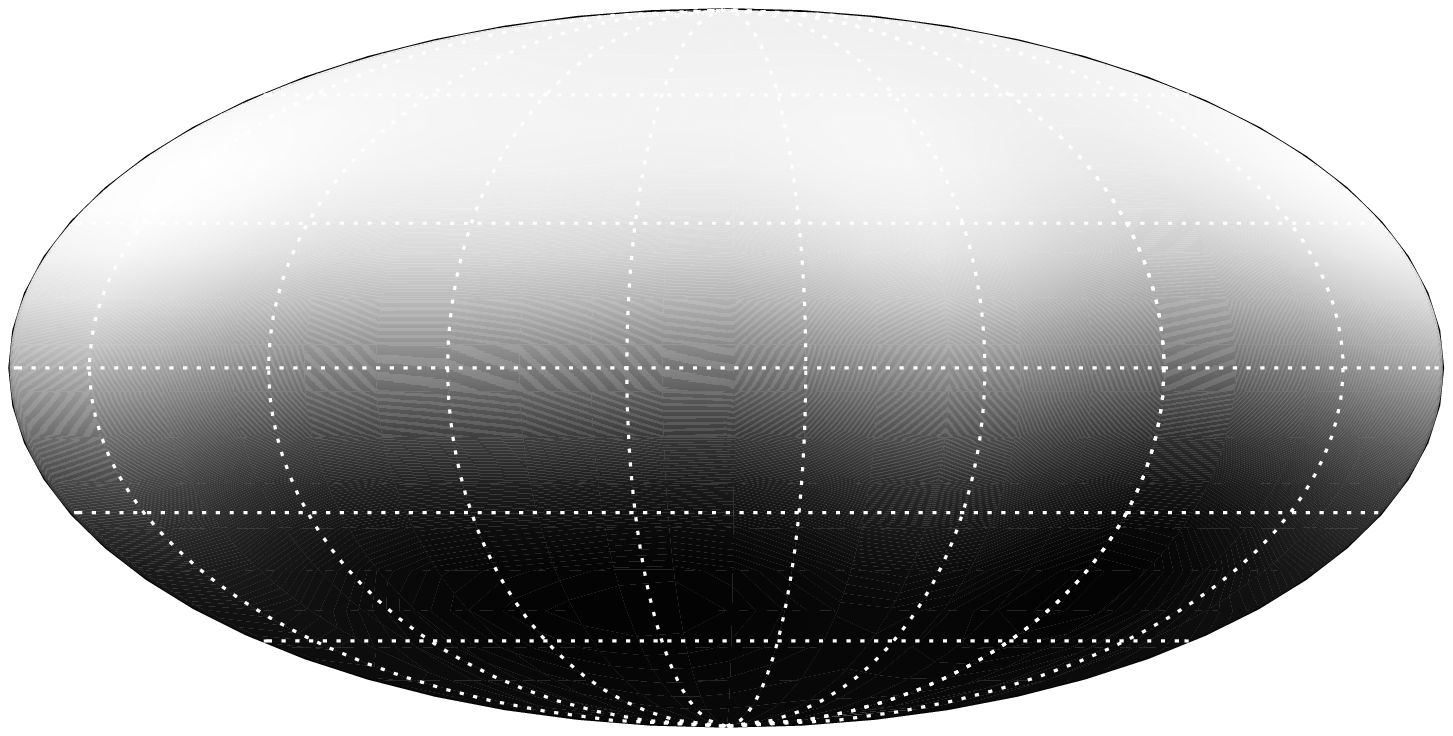,width=9cm}
\vspace*{2mm}
\vskip -7cm \hskip 13cm  {\LARGE 1}
 \vskip 7cm
\vskip -4.5cm
\psfrag{1a}{\Large b}
\hskip 1.5cm \epsfig{figure=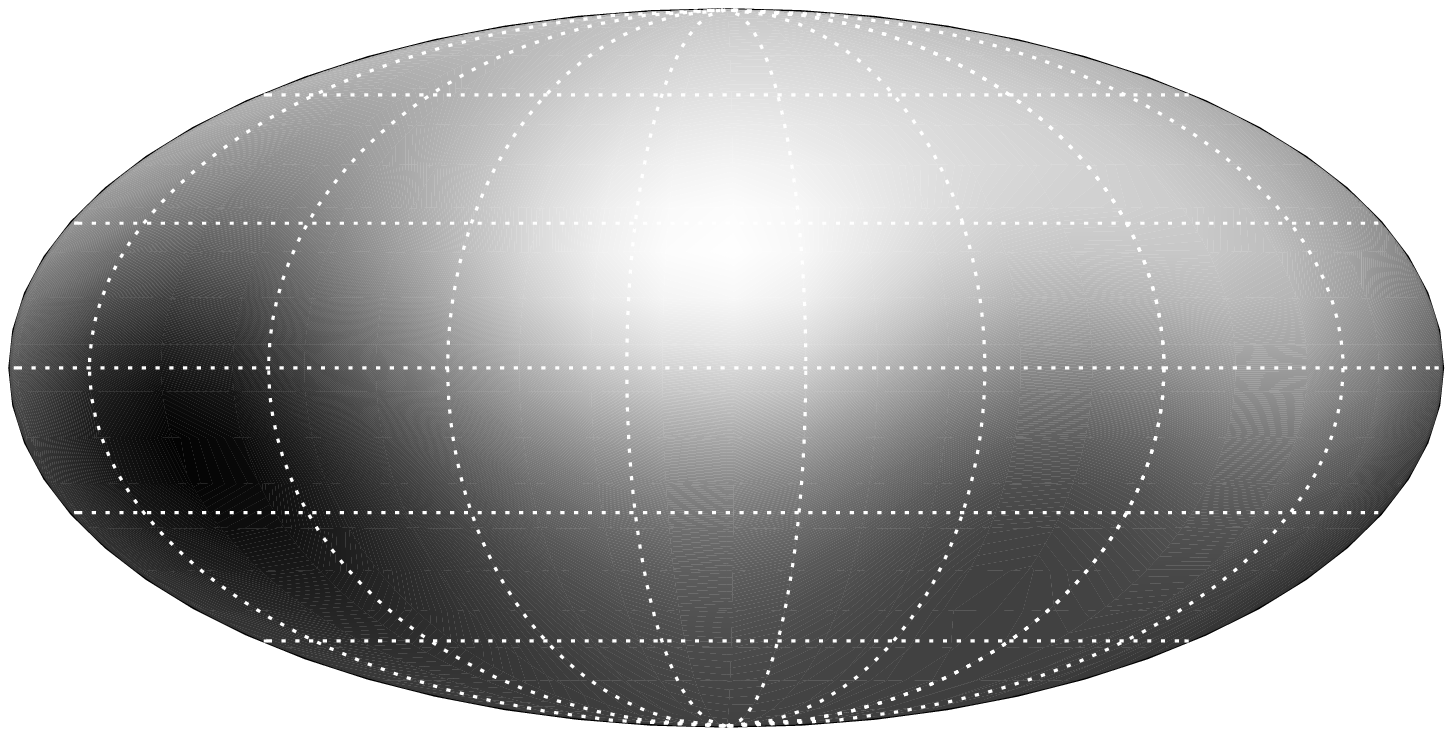,width=9cm}
\vskip -7cm \hskip 13cm  {\LARGE 2}
 \vskip 7cm
\vskip -4.5cm
\psfrag{1a}{\Large c}
\hskip 1.5cm \epsfig{figure=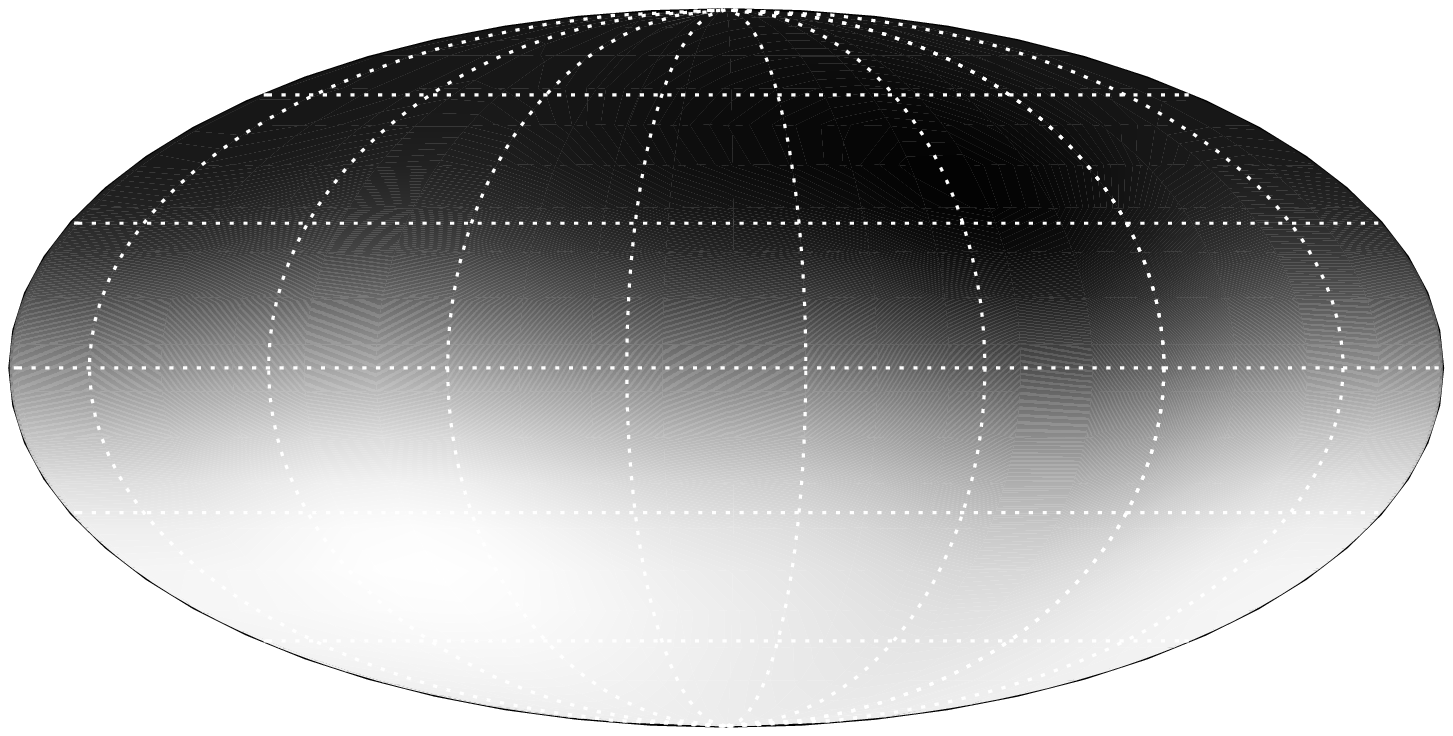,width=9cm}
\vskip -7cm \hskip 13cm  {\LARGE 3}
 \vskip 7cm
%
\vskip -0.4cm
 \caption{The map of  $B_r$-component 
 of the magnetic field in the Mollweide projection for the time moments 
$t=2500$ (1), 2600 (2), 2700 (3). 
 The white colour corresponds to the positive values and the black one to the negative.
}
\label{fig3}
\end{figure}
}

{\Huge
\begin{figure}[th]
\psfrag{Sp}{\Large $g_1^0$}
\psfrag{t}{\Large $t$}
\psfrag{1a}{\LARGE 1}
\hskip -0.0cm \epsfig{figure=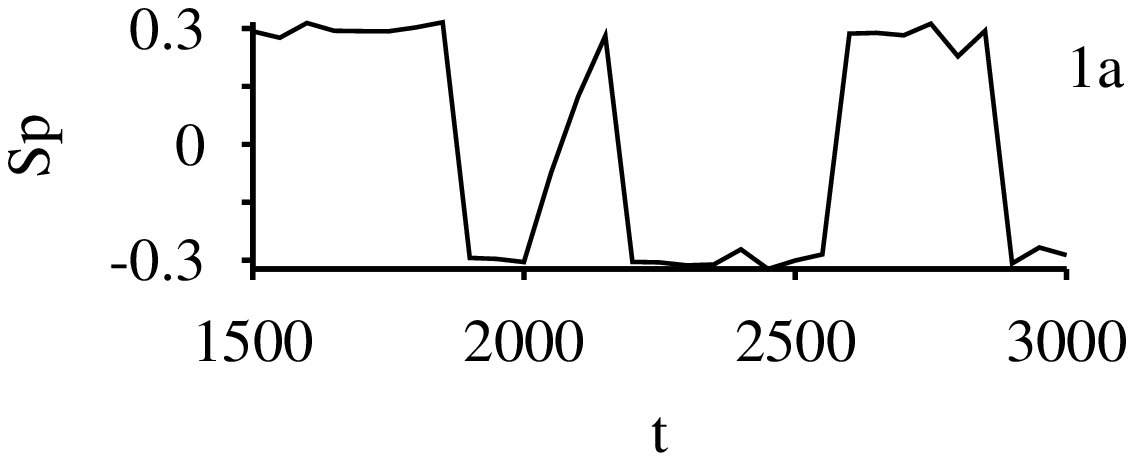,width=13cm}
\vspace*{2mm}
\psfrag{1a}{\LARGE 2}
\psfrag{Sp}{\Large ${\cal D}$, $\cal Q$}
\hskip 2.5cm \epsfig{figure=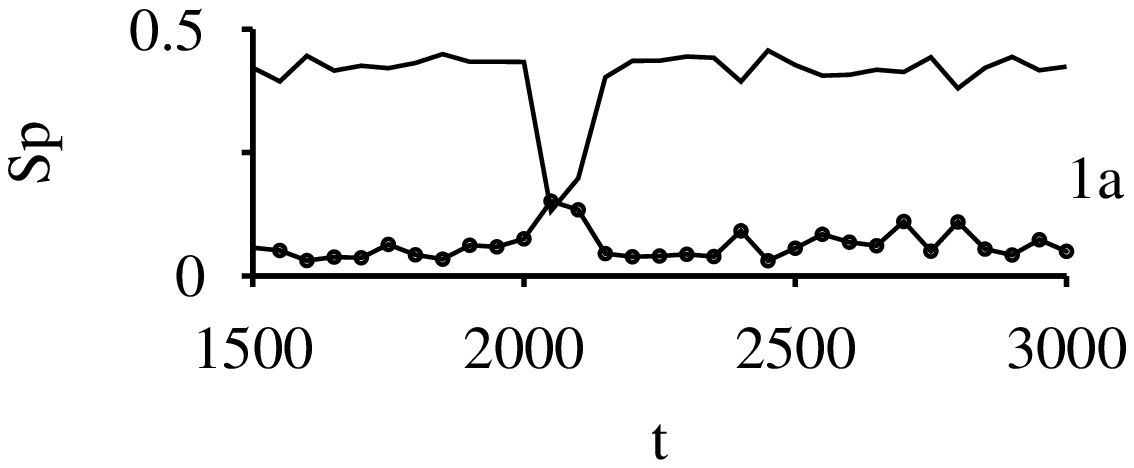,width=13cm}
%
\vskip 0.4cm
 \caption{
Evolution of   $g_1^0$ (1)  and  (2) ${\cal D}=\sqrt{2}\,\sqrt{{g_1^0}^2+{g_1^1}^2+{h_1^1}^2}$ (solid line), 
${\cal Q}=\sqrt{3} \,\sqrt{{g_2^0}^2+{g_2^1}^2+{h_2^1}^2+
+{g_2^2}^2+{h_2^2}^2}$ (circles).
}
\label{fig4}
\end{figure}
}

\eject\newpage

{\Huge
\begin{figure}[th]
\psfrag{t}{ $t$}
\psfrag{M}{$M$}
\psfrag{Vf}{$V_\varphi$}
\psfrag{1a}{\Large 1}
\hskip 0.5cm \epsfig{figure=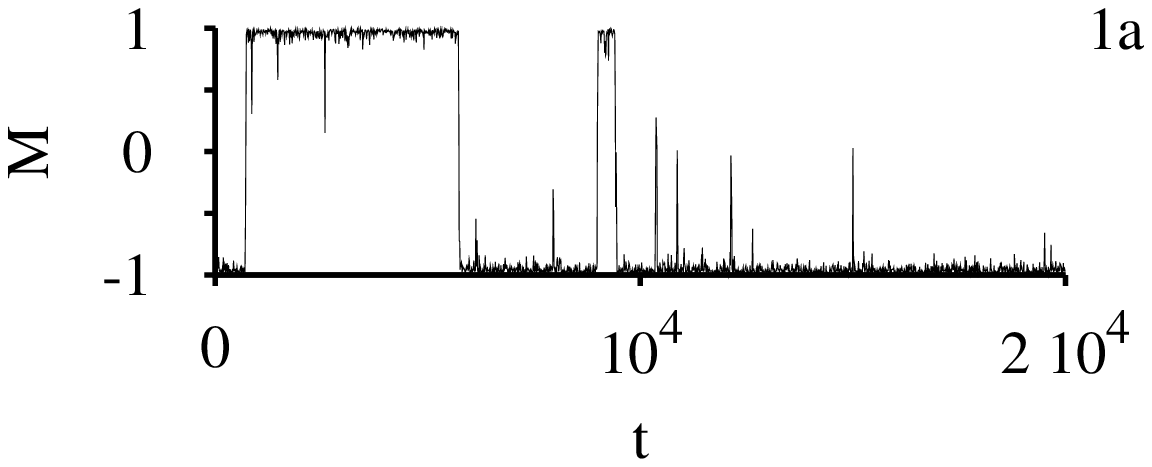,width=9cm}
\vskip 0.1cm
\psfrag{1a}{\Large 2}
\hskip 0.5cm \epsfig{figure=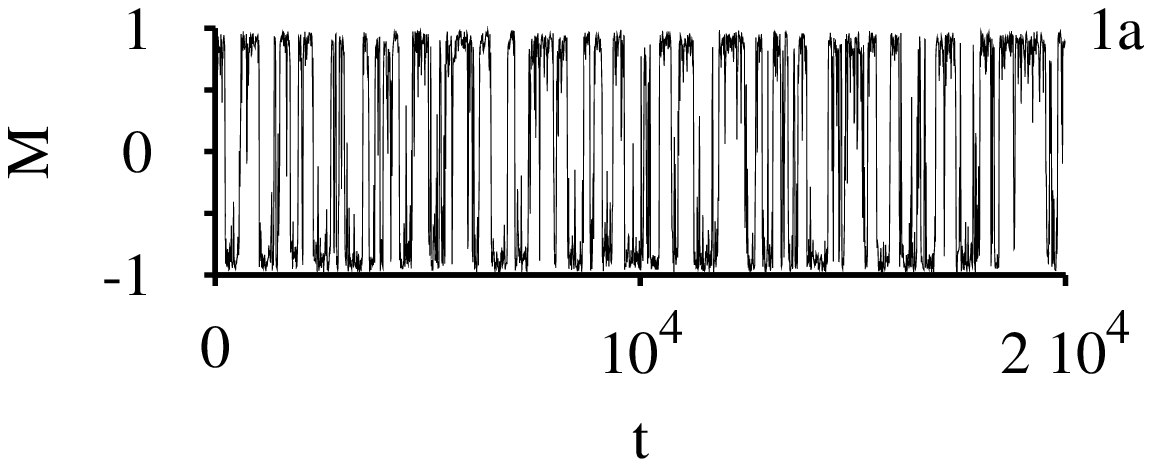,width=9cm}
\vskip 0.1cm
\psfrag{1a}{\Large 3}
\hskip 1.0cm \epsfig{figure=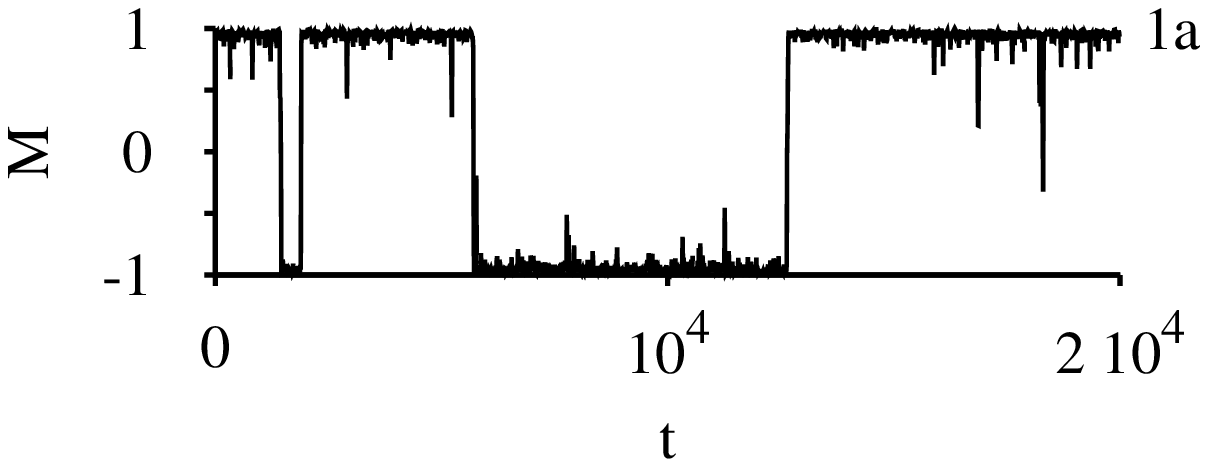,width=9cm}
%
\vskip 0.4cm
 \caption{  Evolution of  $M$ for the regime in Fig.\ref{fig2}a with   $\psi=\cos^2\theta$:
    $C_\psi=2$ (1),   $C_\psi=-1$ (2) and    
  $C_\psi=3$ with   $\psi=\cos^2\theta\cos^2\varphi$ (3).
 }
\label{fig5}
\end{figure}
} 

\eject\newpage

{\Huge
\begin{figure}[th]
\psfrag{h}{\large $h_1^1$}
\psfrag{g}{\large $g_1^1$}
\psfrag{1a}{\Large 1}
\hskip 0.5cm \epsfig{figure=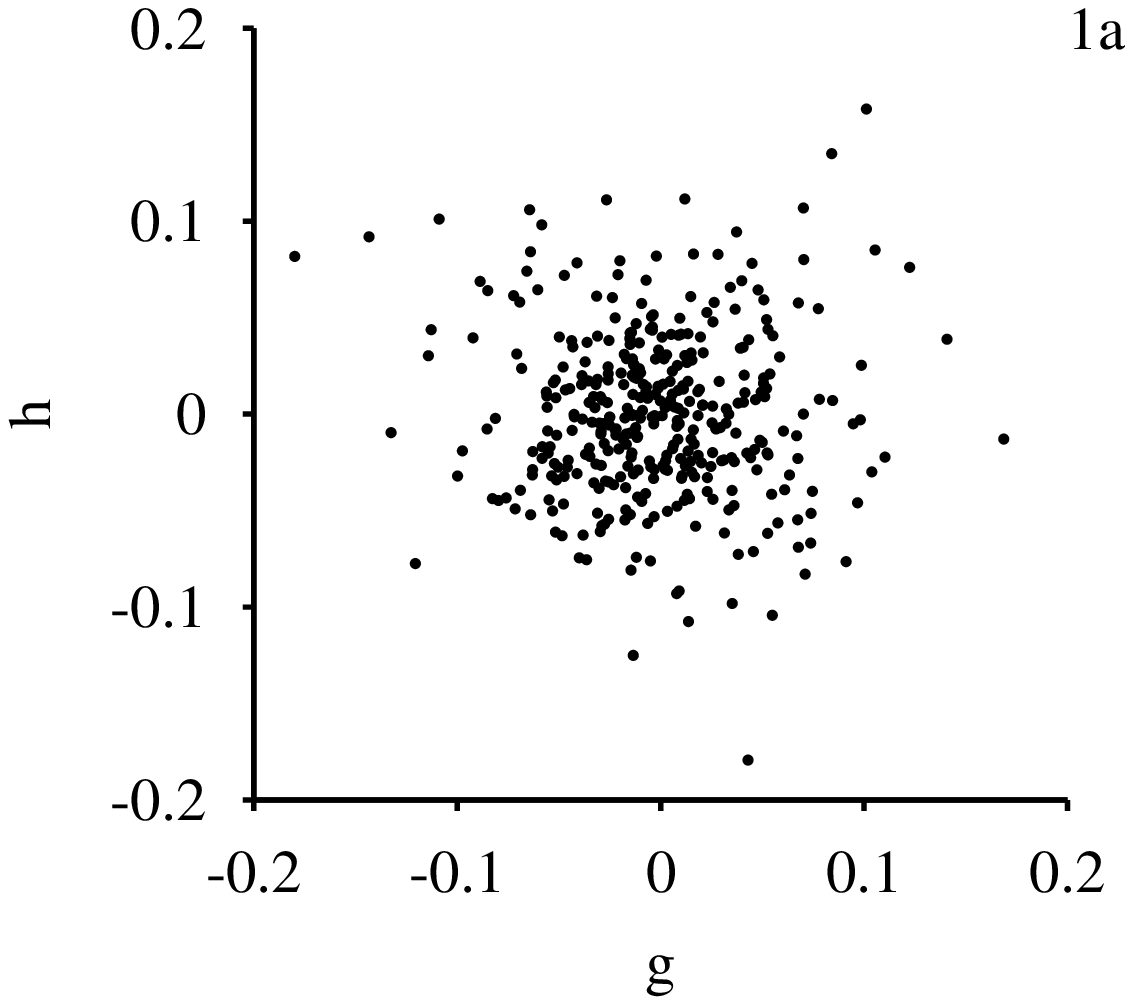,width=9cm}
\vskip 0.1cm
\psfrag{1a}{\Large 2}
\hskip 0.5cm \epsfig{figure=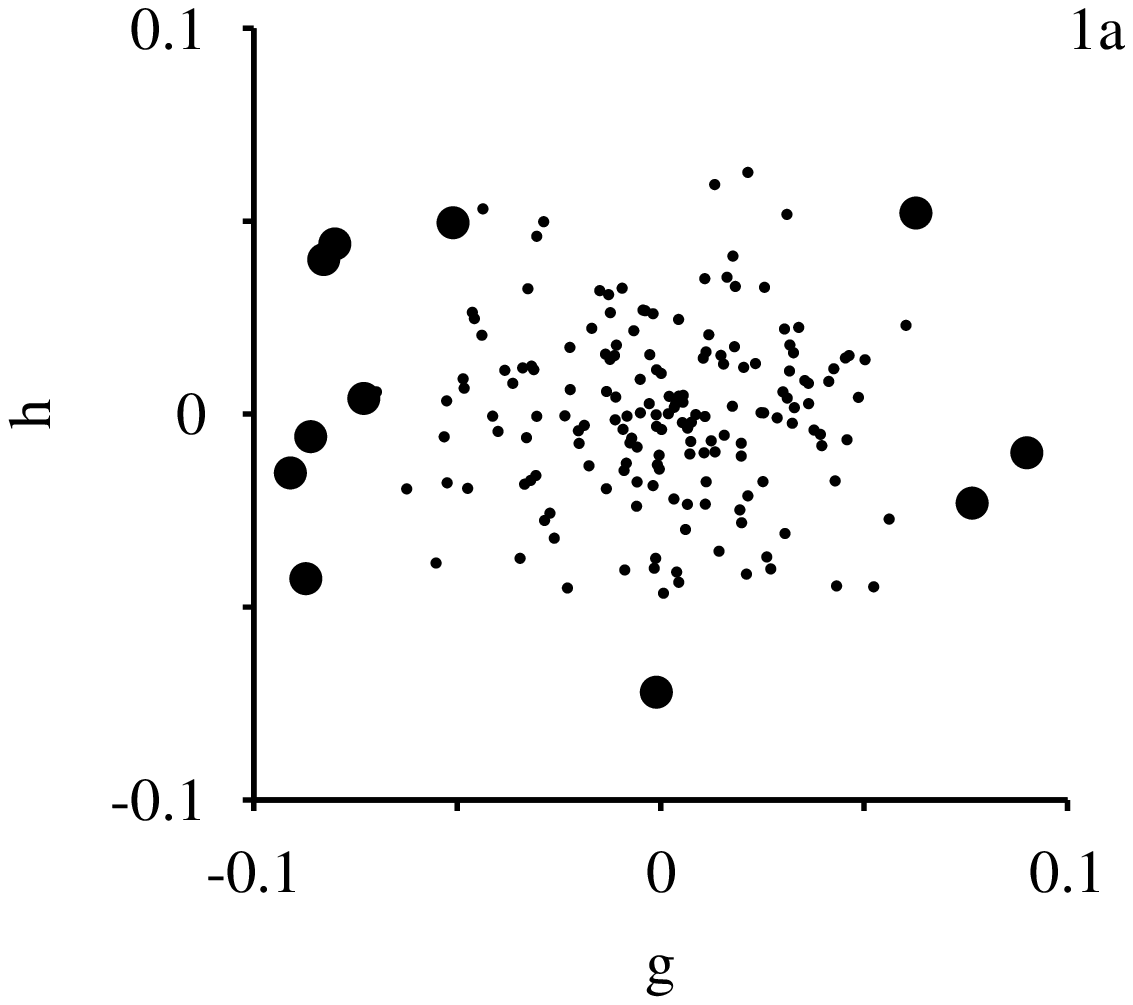,width=9cm}
\vskip 0.1cm
\psfrag{1a}{\Large 3}
\hskip 0.5cm \epsfig{figure=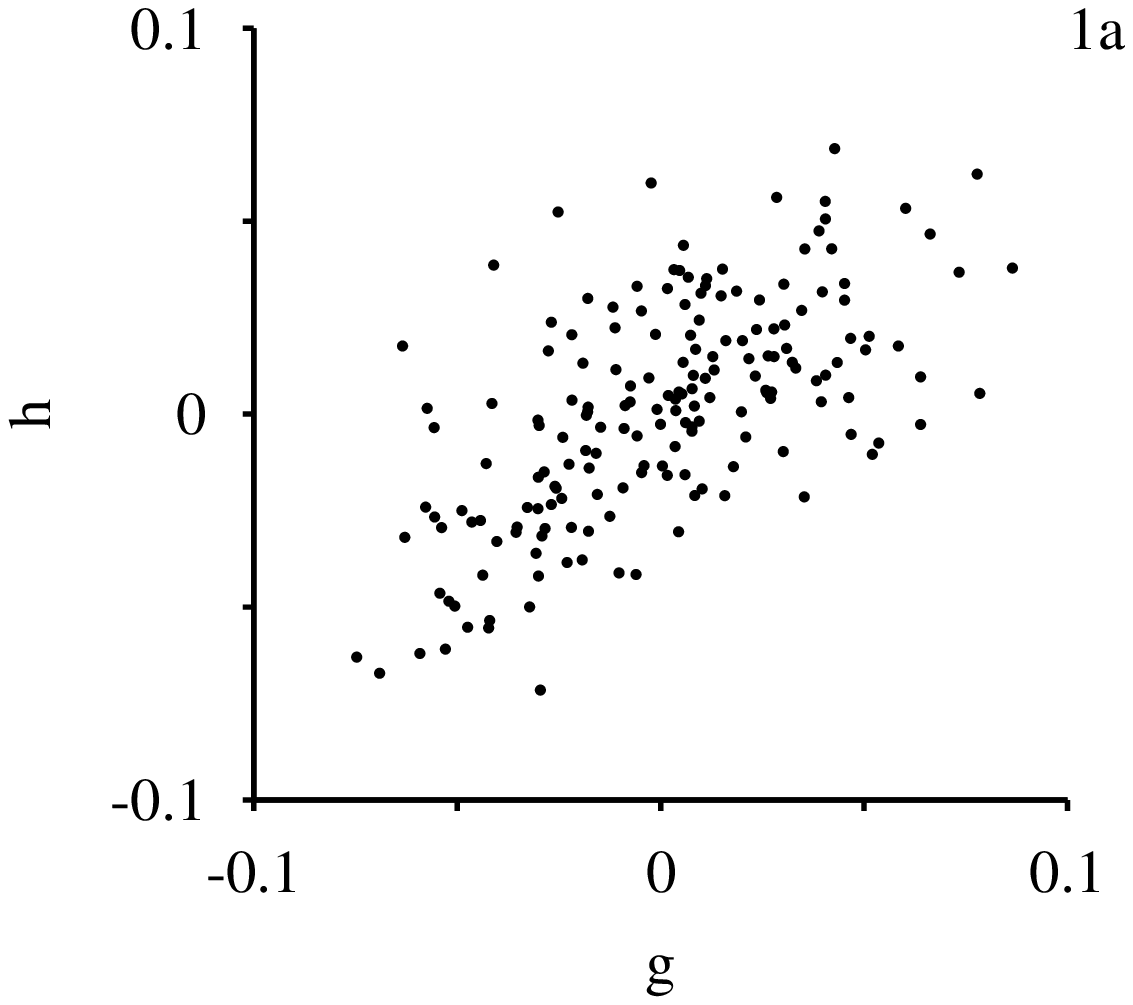,width=9cm}
%
\vskip 0.4cm
 \caption{   Distribution of  $h_1^1(g_1^1)$ for regime in Fig.\ref{fig2}a (1), for regime 
 in Fig.\ref{fig5}a, where  
 large circles correspond to condition  $h_1^{1\, 2}+g_1^{1\, 2}>5\, 10^{-3}$ (2) and 
 for $\psi=\cos^2\theta\, \sin 2\varphi$, $C_\psi=3$ (3).
}
\label{fig6}
\end{figure}
} 

\eject\newpage


\begin{thebibliography}{00}

\bibitem[Cupal  et al.(2002)]{CHR02} 
 Cupal, I., Hejda, P., Reshetnyak, M., 2002. 
Dynamo model with thermal convection and with the free-rotating inner core.  
 Planet.  Space Sci.  50,   P.1117--1122.

\bibitem[Glatzmaier et al.(1999)]{GCHR99} 
Glatzmaier, G.A.,   Coe, R.S.,  Hongre, L.,  Roberts, P.\,H., 1999. 
 The role of the Earth's mantle in controlling the frequency of geomagnetic reversals. 
Nature.  401,  885--890.

 \bibitem[Hejda and Reshetnyak(2009)]{HR09} 
Hejda, P.,  Reshetnyak, M., 2009. Effects of anisotropy in the geostrophic turbulence. 
 Phys. Earth Planet. Int.  177,  152--160.

 \bibitem[Hollerbach(2003)]{Hol} 
Hollerbach, R.,  2003. 
The range of timescales on which the geodynamo operates. 
V. Dehant, K. Creager, S.-I. Karato \& S. Zatman (Editors),
 Earth's Core: Dynamics, Structure, Rotation, AGU Geodynamics Series.  31, 181--192. 

\bibitem[Jacobs(2005)]{Jacobs}
Jacobs, J.A., 2005.  Reversals of the Earth's Magnetic Field.
Cambridge University Press, Cambridge.

\bibitem[Jones(2011)]{Jones}
Jones, C.A., 2011.  Planetary magnetic fields and fluid dynamos. Phil.
Trans. R. Soc. London.    43, 583--614.

\bibitem[Krause and R\"adler(1980)]{Kr}
Krause F., R\"adler K.-H., 1980.  Mean field magnetohydrodynamics
and dynamo theory.    Akademie-Verlag, Berlin.

\bibitem[Landau and Lifshitz(2005)]{Landau}
Landau, L.D.,  Lifshitz, E.M., 2002. Mechanics, 3rd Edition.  Butterworth�Heinemann, Oxford.

\bibitem[Miltat et al.(2002)]{MAT02} 
 Miltat, J., Albuquerque, G., Thiaville, A., 2002. 
An introduction to micromagnetics in the dynamic regime. 
Topics Appl. Phys. 83,  1--34.

\bibitem[Nakamichi et al.(2012)]{NM12} 
Nakamichi, A., Mouri, H.,  Schmitt, D.,  Ferriz-Mas, A.,  Wicht, J.,  
 Morikawa, M., 2012. 
Coupled spin models for magnetic variation of planets and stars. 
  Mon. Not. R. Astron. Soc. 
423, 4, 2977--2990, arXiv:1104.5093v1.

\bibitem[Olson  et al.(2002)]{Olson10} 
 Olson, P.L., Coe, R.S,   Driscoll, P.E., Glatzmaier, G.A, Roberts, P.H., 2010. 
Geodynamo reversal frequency and heterogeneous core--mantle boundary heat flow.
  Phys. Earth Planet. Inter. 180,  66--79. 

\bibitem[Olson(2007)]{Tret} 
Treatise on Geophysics, V.8: Core Dynamics, Olson, P.,
Ed., London: Elsevier, 2007.

\bibitem[Parker(1955)]{Parker55} 
Parker, E.N., 1955.  Hydromagnetic Dynamo Models. Astrophys. J.  
 122,  293--314.

\bibitem[Pedlosky(1987)]{Pedlosky}
Pedlosky J., 1987.  Geophysical Fluid Dynamics.
Springer-Verlag, NY.

\bibitem[Rudiger and   Hollerbach(2004)]{RH04}
Rudiger, G.,   Hollerbach, R., 2004. The Magnetic Universe: Geophysical and Astrophysical Dynamo Theory.  
Wiley VCH, Weinheim.

\bibitem[Shatsillo  et al.(2005)]{Shats05} 
  Shatsillo, A. V.,   Didenko, A.N.  Pavlov, V.E., 2005. 
 Two competing Paleomagnetic directions in the Late Vendian:
  New data for the SW Region of the Siberian Platform.  Russ. J. Earth
 Sci. 7,  4,    3--24.

\bibitem[Stanley(1971)]{Stanley}
Stanley, H.E., 1971. Introduction to phase transitions and critical phenomena. 
Clarendon Press, Oxford. 

\end{thebibliography}
\end{document}